\documentclass[preprint]{aastex}
\usepackage{appendix}
\usepackage{graphicx}
\usepackage{threeparttable}
\usepackage{longtable}
\def\jcap{J. Cosmol. Astropart. P.}
\def\aap{Astron. Astrophys.}
\def\apjs{ApJS}
\def\prd{Phys. Rev. D}
\def\apj{Astrophys. J.}
\def\prl{Phys. Rev. Lett.}

\begin{document}

\title{GeV  observations of the extended pulsar wind nebulae constrain  the pulsar interpretations of the cosmic-ray positron excess}

\author{Shao-Qiang Xi$^{1,3}$, Ruo-Yu Liu$^{2}$, Zhi-Qiu Huang$^{1,3}$, Kun Fang$^{4}$, and Xiang-Yu Wang$^{1,3}$}

\altaffiltext{1}{School of Astronomy and Space Science, Nanjing University, Nanjing 210023, China;
xywang@nju.edu.cn}
\altaffiltext{2}{Deutsches Elektronen-Synchrotron (DESY), Platanenallee 6, D-15738 Zeuthen, Germany; ruoyu.liu@desy.de}
\altaffiltext{3}{Key Laboratory of Particle Astrophysics, Institute of High Energy Physics, Chinese Academy of
Sciences, Beijing 100049, China}
\altaffiltext{4}{Key laboratory of Modern Astronomy and Astrophysics, Nanjing
University, Ministry of Education, Nanjing 210093, China}

\begin{abstract}
{It has long been suggested that nearby pulsars within $\sim 1 \,{\rm kpc}$ are the leading candidate  of the 10-500 GeV cosmic-ray positron excess measured by PAMELA and other experiments. The recent measurement of surface brightness profile of TeV nebulae surrounding Geminga and PSR~B0656+14 by the High-Altitude Water Cherenkov Observatory (HAWC) suggests inefficient diffusion of particles from the sources, giving rise to a debate on the pulsar interpretation of the cosmic-ray positron excess. Here we argue that  GeV  observations provide more direct constraints on the positron density in the TeV nebulae in the energy range of 10-500 GeV and hence on the origin of the observed positron excess. Motivated by this, we search for GeV emission from the TeV nebulae with the \textsl{Fermi} Large Area Telescope (LAT). No spatially-extended GeV emission is detected from these two TeV nebulae in the framework of two-zone diffusion spatial templates, suggesting a relatively low density of GeV electrons/positrons in the TeV nebulae. A joint modelling of the data from HAWC and \textsl{Fermi}-LAT disfavors Geminga and PSR~B0656+14 as the dominant source of the positron excess at $\sim 50-500$ GeV for the usual Kolmogorov-type diffusion, while for   an energy-independent diffusion,  a dominant part of the positron excess contributed by them cannot   be ruled out by the current data.}
\end{abstract}

\section{Introduction}
Measurements of cosmic-ray positron fraction by PAMELA and other experiments have found an excess in the energy range of $\sim 10-500$ GeV\citep{Pamela2009,Fermi2012,AMS2013}, relative to the standard predictions for secondary production in the interstellar medium (ISM). These high-energy positrons should be produced by nearby sources within $\sim {\rm kpc}$ due to severe radiative energy losses in the propagation to Earth (e.g., \citet{Aharonian1995}). The origin of these positrons is still unknown, and the highly suggested candidate sources include nearby pulsars (e.g., \citet{Hooper2009,Yuskel2009,Profumo2011,Yin2013,Razzaque2017}) and  annihilating dark matter particles (e.g., \citet{Cholis2013,Silk2014,Bergstrom2009,Yin2009}). Geminga and PSR~B0656+14 are particularly attractive candidates for $\sim 50-500$ GeV positrons due to their proximity to Earth and suitable ages.

Observations from  Milagro\citep{Milago2009}, along with recent observations by HAWC\citep{Abeysekara2017a},  have revealed extended TeV emissions surrounding  Geminga and PSR~B0656+14. The angular sizes of the extended TeV nebulae are much larger than the X-ray PWNe. The intensity of the TeV emission  indicates that these pulsars deposit a significant fraction of their total spin-down power into high-energy electrons and positrons\citep{Hooper2017}.  Recently, HAWC measured the angular surface brightness profile (SBP) of the TeV nebulae around Geminga and PSR~B0656+14\citep{Abeysekara2017}. The angular SBP suggests that the diffusion coefficient within at least a few tens of pc of these two pulsars is significantly lower than that in the normal ISM. The HAWC collaboration argues that such a low diffusion coefficient leads to a negligible positron flux at Earth, disfavoring them as the sources of the observed positron excess\citep{Abeysekara2017}. However, it was later argued that such a low diffusion region should be only restricted to a small region close to these pulsars\citep{Hooper&Linden2018}.  By invoking a normal diffusion coefficient in the region outside the TeV nebula, the positron flux at Earth can be significantly increased\citep{Fang2018,Profumo2018}. Based on the above arguments, several works  proposed that  Geminga and PSR~B0656+14 remain  the best candidates for producing a dominant fraction of the observed positron excess at $\sim 50-500$ GeV \citep{Hooper&Linden2018,Fang2018,Profumo2018}.

The observed TeV gamma-rays are produced by electrons/positrons with energy above 10\,TeV that up-scatter cosmic microwave background (CMB) photons, according to an approximate relation $E_\gamma=20{\rm TeV}(E_e/100{\rm TeV})^2$, where $E_e$ is the energy of electrons/positrons. Thus, TeV emission does not directly reflect the properties of positrons in the energy range of $10-1000$GeV relevant to the measured cosmic-ray positron excess. Instead, these positrons should produce GeV emission through the inverse-Compton scattering of infrared and optical background photons. If we ascribe the measured positron excess to Geminga and PSR~B0656+14, the required amount of injected electrons/positrons at $50-500$ GeV should be able to produce GeV nebulae around these pulsars. Motivated by this, we first study possible GeV emission associated with the TeV nebulae using the data from \textsl{Fermi}-LAT in \S 2. Then in \S 3, we study whether Geminga and PSR~B0656+14 can be  the dominant source of the positron excess under the GeV constraints. We discuss the influence of various parameters of the diffusion models on the positron flux in \S 4. Finally we give a summary in \S5.

\section{Analysis of Fermi/LAT data}
Previous searches for the GeV emission around Geminga with \textsl{Fermi}-LAT have yielded non-detections\citep{Ackermann2011}. We here analyse the 10-yr  \textsl{Fermi}-LAT data near the region of Geminga and PSR~B0656+14, searching for possible spatially extended emission associated with the TeV nebulae.

\subsection{Count map}
Employing the newest released \textsl{Fermi} Science Tools package  (v11r5p3)  with the instrument response functions (IRFs) P8R3\_SOURCE\_V2, we study the extended gamma-ray emission around Geminga and $\rm PRS\ B0656+14$ using 10 yr (from 2008 August 4 to 2018 September 17) of Pass 8 SOURCE data at energies between 10 GeV and 500 GeV.  We consider the photons within $40^\circ \times 40^\circ$ region of interest (ROI) centered at positions ($l,b$)= ($195.14^\circ, 4.27^\circ$). Photons detected at zenith angles larger than $105^\circ$ were excised to limit the contamination from gamma-rays generated by cosmic-ray interactions in the upper layers of the atmosphere, and were further filtered by the relational filter expression $\rm(DATA\_QUAL>0)\ \&\&\ (LAT\_CONFIG==1)$ in \textsl{gtmktime} tool. We binned our data in 12 logarithmically spaced bins in energy and use a spatial binning of $0.05^\circ$ per pixel.

The background model in this analysis includes: (1) the diffuse Galactic interstellar emission (IEM) template ( $\rm gll\_iem\_v6.fits$) released and described by the \textsl{Fermi}-LAT collaboration through the \textsl{Fermi} Science Support Center (FSSC) \citep{2016ApJS..223...26A}, (2) the isotropic diffuse component accounting for the extragalactic diffuse gamma-ray background and misclassified cosmic rays (CRs) with a spectral shape described by $\rm iso\_P8R3\_SOURCE\_V2.txt$  and (3) the point and extended sources listed in the third Catalog of Hard \textsl{Fermi}-LAT Sources (3FHL) \citep{2017ApJS..232...18A}. In addition, we search for  new gamma-ray point sources using the \textsl{gttsmap} tool and 30 non-3FHL sources are found, which are included in our background model.  The counts map with the point and extended sources of the background model is shown in the left panel of Figure~\ref{fig_cs}.
We first do the background-only fitting. All the spectral parameters of the point and extended sources of the background model are left free. We also left three degrees of freedom to the diffuse emission, i.e., the normalizations of the isotropic and Galactic components, and a spectral index $\rm \Gamma$ that can make the Galactic component a little harder or softer after it is multiplied by $\rm (E/E_0)^{\Gamma}$. Based on the best-fit value from the background-only model fit, we obtain the residual map shown in the right panel of Figure~\ref{fig_cs}. We do not find any obvious excess around the Geminga and the $\rm PSR\ B0656+14$ in the residual map.

\subsection{Two-zone diffusion templates for Fermi/LAT data analysis}
It has been argued that the assumption of a  low diffusion coefficient in the local region is inconsistent with the detection of the highest-energy (up to 20 TeV) electrons by H.E.S.S. \citep{Hooper&Linden2018}. Thus, it is more reasonable to consider that the low diffusion region is only restricted to the  region close to pulsars and the outer region has a normal diffusion coefficient,
as already suggested in \citet{Fang2018} and \citet{Profumo2018}. In addition, possible mechanisms for the suppression of the diffusion coefficient in the region close to pulsars have been suggested\citep{Evoli2018}. Following this, we consider a two-zone model for electron diffusion and adopt a step function for the diffusion coefficient
\begin{equation}
D(E_e,r)=\left\{
\begin{array}{ll}
D_1, \quad r<r_0\\
D_2, \quad r\geq r_0 .
\end{array}
\right.
\end{equation}
Following \citet{Abeysekara2017}, $D_1=4.5\times 10^{27}\left(\frac{E_e}{100\,\rm TeV}\right)^{1/3}{\rm \,cm^2 s^{-1}}$ is considered for  the vicinity of the pulsar, where the Kolmogorov-type diffusion (i.e., with energy dependence $\delta=1/3$) is assumed  for the inner region. For the diffusion coefficient in the ISM, we take the GALPROP default value, $D_2=1.8\times 10^{30}\left(\frac{E_e}{100\,\rm TeV}\right)^{1/3}{\rm \,cm^2 s^{-1}}$,  as inferred from measurements of the boron-to-carbon ratio and other cosmic-ray secondary-to-primary ratios  \citep{Hooper&Linden2018, Profumo2018}. {For convenience, we call such a combination of parameter values (i.e., $r_0=50\,$pc, $D_1=4.5\times 10^{27}\left(\frac{E_e}{100\,\rm TeV}\right)^{1/3}{\rm \,cm^2 s^{-1}}$, $B_1=3\mu$G, $p=2.25$ and $D_2=1.8\times 10^{30}\left(\frac{E_e}{100\,\rm TeV}\right)^{1/3}{\rm \,cm^2 s^{-1}}$) the benchmark case.}  {\citet{Hooper2017}
has shown that, even if all pulsars have such a low-diffusion halo,  there is little impact on the observed secondary-to-primary ratios (e.g., the boron-to-carbon ratio) in the cosmic-ray spectrum as measured at Earth as long as
$r_0\le 50 {\rm pc}$. This is because  such low-diffusion regions occupy only a fraction of $<3\%$ of the volume of the Milky Way¡¯s disk.}

\subsection{Flux limits and TS values for various spatial templates}

We evaluate the test statistic (TS) for an additional extended source, defined as ${\rm TS}=-2({\rm ln}L_0-{\rm ln}L)$, where $L_0$ is the maximum-likelihood value for null hypothesis  and $L$ is the maximum likelihood with the additional extended source described by the templates. First, we create a $2^\circ$ disk template, naively assuming that GeV emission has the same size as that of the TeV halo  in the HAWC catalog\citep{Abeysekara2017a}. Then, we consider a one-zone diffusion template (named Diffusion 1 template hereafter),  assuming that  GeV  emission follows that predicted by the one-zone diffusion model adopting the same parameters as used in \citet{Abeysekara2017}.

Next, we focus on the two-zone diffusion models. We first build a two-zone template considering an inner region as that for TeV halo and an outer region assuming a normal diffusion coefficient in the ISM. We name this template as Diffusion 2 template.

Since the GeV gamma-ray flux overshoots the upper limits of Fermi-LAT for the Diffusion 2 template (see \S3.2 for details), we adjust the electron spectral index from soft to hard and thus obtain the Diffusion 3 template for which the expected flux just touches  the upper limit. In addition, on the premise of fitting the flux and the SBP of the TeV nebulae measured by HAWC, {we create two more extended source templates: i.e., Diffusion 4 template corresponds to the case changing $r_0$=50 pc to 100 pc and Diffusion 5 template corresponds to the case changing the energy dependence of the diffusion coefficient of the inner region to $\delta=0$.} For the GeV emission around the PSR~B0656+14, we simply consider the two-zone diffusion models for a Kolmogorov-type diffusion of inner region (Diffusion 6) and for an energy-independent diffusion of inner region (Diffusion 7). Note that we relocate the ROI center at the position of the PSR~B0656+14 when fitting the Diffusion 6 and Diffusion 7 templates.

{We obtain the radial surface brightness profiles by integrating the gamma-ray flux  over the line of sight for a given angle $\theta$ from the location of the pulsar. The obtained intensity as a function of $\theta$, i.e., $I(\epsilon, \theta)$ is used as the spatial template for analysing the {Fermi-LAT data}. The surface brightness profiles at 10 GeV and  500 GeV are shown as examples in Fig.\ref{fig:SBPGeV}. The angular extension for the GeV emission is much larger than that of the TeV emission since the GeV-emitting electrons can travel a larger distance before losing their energy than TeV-emitting electrons, especially in the one diffusion zone model. In the presence of two diffusion zones, the negative gradient of electron density is large near the boundary between the two diffusion zones, leading to a quick drop of the surface brightness with the distance to the pulsar.

Since the expected spatial distributions of the GeV emission are energy dependent for the above diffusion models,  we create the templates using the \textsl{mapcube} file, which is a 3 dimensional FITS map allowing arbitrary spectral variation as a function of sky position. {We cut the templates at $25^{\circ}$, within which more than $90\%$ of the integrated flux lie for all the above diffusion templates.} Note that  the spectral parameters of the background sources are still left free  for the analysis of each template.

No significant extended GeV emission is detected from the TeV nebulae of Geminga and PSR~B0656+14. The  flux limits for various spatial templates are obtained and are shown in Table~\ref{tab1}. Spatial templates with larger angular extension usually result in larger TS values. {The large difference in the limits  for different templates is due to the large difference in the sizes and angular profiles of these templates}.

To assess the robustness of our results listed in Table~\ref{tab1}, we perform a number of systematic checks such as varying the ROI size and selecting different event class data. In particular, we investigate how the diffuse Galactic emission modelling affects the TS values and the upper limits using eight alternative IEM templates\citep{2012ApJ...750....3A,2016ApJS..224....8A}. {The details of these eight alternative IEM templates are given in the Appendix.  The total systematic uncertainties of the upper limits for the diffusion templates are about $32\%$ (see Table 2 in the Appendix).}

\section{Implications for the positron flux at Earth}
Below we argue that the flux limits of gamma-rays in $10-500$ GeV  can  impose meaningful constraints on the positron flux nearby the pulsars and hence on their contribution to the positron flux observed at Earth.

\subsection{The electron/positron flux calculation }
To obtain the theoretical GeV emission from the TeV nebulae, we first need to calculate the distribution of electrons/positrons, which can be obtained through the following transport equation, assuming the pulsar is located at $r=0$,
\begin{equation}
\frac{\partial N(\gamma,r,t)}{\partial t}= \frac{1}{r^2}\frac{\partial}{\partial r}\left(r^2D(\gamma,r)\frac{\partial N}{\partial r} \right) -\frac{\partial}{\partial \gamma}\left(\dot{\gamma}N\right)+Q(\gamma, t)\delta(r),
\end{equation}
where $N(\gamma,r,t)$ represents the  density of electrons/positrons with Lorentz factor $\gamma$ at the space time of $(r, t)$, $D(\gamma,r)$ is the diffusion coefficient at a distance $r$ to the pulsar, $\dot{\gamma}$ is the cooling rate of electrons/positrons due to both synchrotron radiation and inverse Compton radiation, and $\delta(r)$ is the Dirac delta function which signifies electron/positron injection only from the pulsar.
The cooling rate taking into account the Klein-Nishina effects can be approximately given by \citet{Moderski2005}
\begin{equation}
\dot{\gamma}=\frac{4\sigma_T\gamma^2}{m_ec}\left[U_B+{\sum}_{i} \frac{U_{\rm ph, i}}{(1+4\gamma \frac{\epsilon_{T,i}}{m_ec^2})^{3/2}} \right],
\end{equation}
where $U_B$ is the magnetic field energy density and $U_{\rm ph, i}$ represents the radiation field energy density, including CMB, infrared and optical photons. The photon fields are considered to have a grey body distribution, for which the temperature and energy density are approximately those derived by GALPROP\citep{Galprop1998,Abeysekara2017}. $\epsilon_{T,i}$ is the average photon energy of the radiation field which is equal to $2.82kT$ in the case of black body or grey body radiation with $k$ being the Boltzmann constant and $T$ being the temperature. The injection term is assumed to be $Q(\gamma,t)=J(\gamma)S(t)$, where $J(\gamma)\propto \gamma^{-p}e^{-\gamma/\gamma_{\rm max}}$ with $p$ being the spectral index and $\gamma_{\rm max}$ being the maximum Lorentz factor of electrons injected by the pulsar, and $S(t)\propto 1/(1+t/\tau)^2$ assuming that the pulsar is a pure dipole radiator of a braking index of 3 {and the spin-down timescale of the pulsar $\tau=12000$yr}. The normalization of the injection term can be found by $\int\int Q (\gamma,t)\gamma m_ec^2d\gamma dt=W_e$, where $W_e$ is the total injection power of electrons and positrons. The numbers of injected electrons and positrons are the same at any energy.
We solve the equation by discretizing the equation with a numerical method similar to that employed in \citet{Fang2018}.

\subsection{One-zone diffusion scenario}
First, we reproduce the gamma-ray flux and the SBP in $8-40\,$ TeV  of the TeV nebula of Geminga measured by HAWC\footnote{except that the injection electron/positron power $W_e$ in our calculation is about 0.5 times of that in \citet{Abeysekara2017}}, by adopting the same parameters used in \citet{Abeysekara2017}, which assumes that the inefficient diffusion region inferred from TeV nebulae extends all the way to Earth. As we can see from the left panel of Fig.~\ref{fig:fitting_hawc}, the contribution to the positron flux by Geminga is very low due to the very low diffusion coefficient, consistent with the result in \citet{Abeysekara2017}. However, while HAWC's measurement is well fitted, the expected GeV fluxes exceed the upper limits of \textsl{Fermi}-LAT by about one order of magnitude. This demonstrates the importance of considering the GeV observations consistently. By invoking a hard electron spectrum, one can in principle get a consistently low GeV flux. Since our work focus on the two-zone diffusion scenarios, we will not discuss this in detail.

\subsection{Two-zone diffusion scenario}
As  shown in Fig.~\ref{fig:Diffusion2}, the multi-TeV spectrum and the SBP can be fitted with most parameters remaining unchanged except introducing an additional outer fast-diffusion region at $r>r_0=50\,$pc, while the predicted positron flux at Earth is significantly increased, accounting for 70\% of the measured positron flux above 100\,GeV. This is consistent with the conclusion in previous literature {\citep{Fang2018, Profumo2018, Tang2018}}. However, the expected GeV gamma-ray flux also overshoots the upper limits of \textsl{Fermi}-LAT by more than one order of magnitude. This demonstrates the important role of the GeV observations in constraining the positron flux from pulsars.

On the premise of fitting the flux and the SBP of the TeV nebulae measured by HAWC, we need to reduce the ratio between  multi-GeV gamma-ray flux and multi-TeV gamma-ray flux. As a result, a harder electron injection spectrum is required or a spectral break needs to be introduced around TeV in order to reconcile the predicted multi-GeV flux with the observed flux by \textsl{Fermi}-LAT. We here adopt the former choice since the number of the free parameters in this case is less. Note that although \cite{Abeysekara2017} suggests a spectral index of $p\simeq 2.25$ to fit the multi-TeV spectrum, a harder electron injection spectrum in combination with a relatively small $\gamma_{\rm max}$ can lead to a similar spectral shape in the range of $8-40\,$TeV.

There are various sets of parameters that are able to fit the multi-TeV spectrum and the SBP, but the corresponding positron fluxes at Earth are different. We need to find out the maximally allowed positron flux at  Earth produced by Geminga and in the mean time take into account the observations of HAWC and \textsl{Fermi}-LAT. Given that the diffusion
in the inner region generally dominates the propagation time of positrons from the pulsar to Earth, we require
\begin{equation}\label{eq:D-r0}
D_1(100\,{\rm GeV})\gtrsim r_0^2/4\tau_{\rm Gem}=5\times 10^{26}\left(\frac{r_0}{50\,{\rm pc}} \right)^2\,\rm cm^2 s^{-1},
\end{equation}
since otherwise the injected positrons have not yet arrived at Earth after a propagation time equal to  the lifetime of Geminga ($\tau_{\rm Gem}=340\,$kyr). On the other hand, if the diffusion coefficient is too large, the injected electrons/positrons will distribute in a very large volume and hence the flux at Earth will also be low. Furthermore, given the fast drop in the HAWC measured SBP within $5^\circ$, the spatial extension of multi-TeV-emitting electrons/positrons should not significantly exceed 25\,pc. Since the diffusion distance of these very high energy electrons/positrons is determined by the radiative loss, we find
\begin{equation}\label{eq:D-B1}
D_1(100\,{\rm TeV})\lesssim 3\times 10^{27}\left[\left(\frac{B_1}{3\mu G}\right)^2+0.6\right]\, \rm cm^2 s^{-1}.
\end{equation}
where $B_1$ is the magnetic field of the inner diffusion region and the coefficient $0.6$ in the square bracket accounts for the IC cooling on CMB photons. For the considered Kolmogorov-type diffusion  (i.e., $D\propto E^{1/3}$), we find that $D_1({E_e})=5\times 10^{26}(E_e/100\rm GeV)^{1/3}\,cm^2s^{-1}$ satisfies the above two equations for  $r_0=50\,$pc and $B_1=3\mu$G. We show the expected gamma-ray spectrum, SBP and positron flux at Earth for such a set of parameters in Fig.~\ref{fig:2D}. Such a set of parameters is very close to our benchmark case, except that a hard injection spectrum with $p=1.7$ (or harder) is needed in order not to overshoot the upper limits imposed by \textsl{Fermi}-LAT. The resulting maximum positron fluxes above $\sim 100$ GeV is less than 10\% of the measured value by AMS-02 \citep{Aguilar14}.

{Based on Eq.~(\ref{eq:D-r0}),} we expect that the positron flux at Earth increases with the size of the inner inefficient diffusion region $r_0$. This is because a larger $r_0$ corresponds to a larger $D_1$ and consequently a faster diffusion of particles. As a result, it requires a larger electron/positron injection power (i.e., $W_e$) to fit the TeV gamma-ray flux, subsequently leading to a higher positron flux at Earth. However, a larger $D_1$ also results in a flatter SBP. To make the predicted SBP as steep as the measured one, we need to invoke a relatively high magnetic field ($B_1$) for the inner region { as indicated by Eq.~\ref{eq:D-B1}. We now consider a large size for the inner zone with $r_0=100\,$pc,  requiring $D_1(100\,{\rm GeV})\gtrsim 2\times 10^{28}\rm cm^2s^{-1}$ according to Eq.~(\ref{eq:D-r0}). The magnetic field of the inner region $B_1>8\mu$G should also be invoked according to Eq.~(\ref{eq:D-B1}). The required magnetic field is larger than the typical ISM magnetic field (i.e., $3-6\,\mu$G), and together with the large $r_0$ leads to a too large magnetic field energy in the inner region, i.e., $W_{B,1}=B_1^2r_0^3/6\simeq 3\times 10^{50}\rm \,$ergs, which far exceeds the total energy of injected electrons/positrons from pulsar spin-down and is comparable to the total energy released in a supernova explosion. One has to seek a very efficient mechanism for amplification of the magnetic field with the energy being provided by some extra sources other than pulsars, or alternatively argue that the high magnetic field  somehow originally pre-existed in this region. Either of the two choices requires extreme conditions. }

\section{Influence of some parameters}
In the above text, we have already discussed the influence of $r_0$, $D_1$ and $B_1$ on the result. We here discuss the influence of the energy dependence of $D_1(E_e)$ (i.e., $\delta$) and the spin-down timescale of Geminga $\tau$.

\subsubsection{ Energy dependence of the diffusion coefficient in the inner region $D_1$}\label{sec:delta}
If we relax the pre-assumption of the Kolmogorov-type diffusion coefficient for the inner region, we can get a relatively large $D(100\,\rm GeV)$ while keeping the value of $D(100\,\rm TeV)$ the same by assuming a weaker rigidity dependence (a smaller $\delta$) in the diffusion coefficient, which subsequently avoids the requirement of a too large magnetic field. Let us take an energy-independent diffusion coefficient (i.e., $\delta=0$) for the inner region as an example. We normalize $D_1$ to $4.5\times 10^{27}\rm cm^2s^{-1}$ at 100\,TeV, which is the best-fit diffusion coefficient from HAWC's observation. The energy-independence of $D_1$ increases the diffusion coefficient at 100\,GeV by one order of magnitude compared to the benchmark case. As a result, the density of GeV-emitting electrons/positrons residing in the inner region is significantly reduced, and the resulting multi-GeV flux is reduced to a level consistent with the \textsl{Fermi}-LAT's upper limits (see Fig.~\ref{fig:constant-D1}) for a softer electron index of $p=2.25$. On the other hand, the faster diffusion leads to a more extended spatial distribution of $\sim 100\,$GeV positrons and the positron flux at Earth reaches a level of $30\%$ of the measured flux above $100$\,GeV. In this case, if we further tune down the employed diffusion coefficient for the outer region $D_2$ by a factor of 2-3, the positron flux can be increased to a level comparable to the measured flux.

For a more extended exploration on the influence of $\delta$, we fix $D_1(100\rm TeV)=4.5\times 10^{27}\rm cm^2s^{-1}$ and adjust $\delta$ to see the change. In Fig.~\ref{fig:delta}, we compare the expected gamma-ray fluxes (top panel), SBPs in $8-40\,$TeV (middle panel) and the spatial distributions of $100$\,GeV positron energy density (bottom panel) for $\delta=0$, 1/10, 1/5, 1/3, 1/2 and 1. All the other parameters are the same as those in the benchmark case. We can see that while the flux and the SBP measured by HAWC are fitted, the GeV fluxes and the distributions of positron density with different $\delta$ vary with $\delta$ significantly. Except for the case of $\delta=0$, the expected GeV fluxes overshoot the \textsl{Fermi}-LAT's upper limits. This is understandable: since the diffusion coefficient at 100\,TeV is fixed, a larger $\delta$ results in a smaller diffusion coefficient at $100-1000\,$GeV that are responsible for the multi-GeV flux. Consequently, lots of GeV-emitting electrons/positrons reside in the vicinity of Geminga and give rise to relatively high GeV fluxes. We can also see that $\delta=1/3$ results in the highest positron flux at 100\,GeV among all the considered $\delta$. It can be understood as follows: given $D_1(100\,\rm TeV)=4.5\times 10^{27}\rm cm^2s^{-1}$ and $\delta=1/3$, the resulting diffusion coefficient for 100\,GeV electrons/positrons satisfy the relation shown by Eq.~\ref{eq:D-r0}. A larger $\delta$ results in a smaller diffusion coefficient for 100\,GeV electrons/positrons so that the amount of them reaches Earth is low. On the contrary, a small $\delta$ results in a larger diffusion coefficient and makes the electrons/positrons distribute in a more extended volume, leading to a lower positron flux  at  Earth when the total injection is fixed. Lastly, we note that the energy-independent diffusion is quite different from the convection-dominated transportation. In convection-dominated transportation model, the CR density would drop much faster than that in the diffusion model, so this model has difficulty in explaining the measured SBP.

\subsubsection{The spin-down timescale $\tau$}\label{sec:tau}
{The spin-down timescale of a pulsar depends on the the surface magnetic field  and the initial rotation period of the pulsar. In all the above calculations, we have taken $\tau=12000$yr following ref.\cite{Abeysekara2017}. However, the uncertainties in the above properties of Geminga can easily lead to a difference of a factor of a few  in the spin-down timescale. Note that TeV-emitting electrons/positrons cool quite fast with a cooling time of $10^{11}$s, which is only about 1\% of the Geminga's age. Thus, the observed TeV emission is mainly contributed by electrons/positrons injected very recently. A larger spin-down timescale means a higher injection power at late time provided that the total injection energy is the same. Thus, we can use a smaller total injection power to fit HAWC's observation, and this would also decrease the multi-GeV gamma-ray flux and the 100\,GeV positron flux. In contrast, a smaller spin-down timescale leads to a smaller injection power at late time and a larger total injection energy is needed to fit HAWC's observation. In this case, the multi-GeV flux and the 100\,GeV positron flux will be increased. In the Fig.~\ref{fig:tau}, we show the results with $\tau=12000\,$yr, 24000\,yr, 6000\,yr and 1000\,yr, while other parameters are the same as those in the benchmark case. By comparing the results in former three cases, we can see that the positron flux at  Earth does not change much by varying $\tau$ by a factor of 2. Particularly, the positron flux in the case of $\tau=24000\,$yr is quite low while the multi-GeV flux almost saturates the \textsl{Fermi}-LAT's upper limit, implying that the positron flux cannot be much increased by employing a softer injection spectrum. Based on this result, we can also expect that an even larger $\tau$ would only result in a lower positron flux. In accordance with our analysis above, $\tau=6000$yr results in a little higher positron flux, but still much lower than the measured one. This motivates us to look into the result with an even smaller spin-down timescale of $\tau=1000$yr.  As we can see, although the positron flux is enhanced with such a small $\tau$, the multi-GeV flux significantly exceeds the \textsl{Fermi}-LAT's upper limit. We thus conclude that the uncertainty in the spin-down timescale will not affect our conclusion.

\subsection{The contribution to the positron flux by PSR~B0656+14}
{PSR~B0656+14 is another nearby pulsar that is suggested to be able to contribute significantly to the positron flux.  The positron flux contributed by  PSR~B0656+14 can be calculated in the same way. The difference is the age of PSR~B0656+14 (110~kyr) and the distance to Earth (288\,pc). According to Eq.~1, the maximum positron flux is obtained when $D_1(100{\rm GeV})=1.7\times10^{27} {\rm cm^2 s^{-1}}$ for $r_0=50 {\rm pc}$. To fit the spectral and SBP data of PSR~B0656+14 and simultaneously reconcile the upper limits imposed by \textsl{Fermi}-LAT, we find that a spectral index  harder than $p=-1.9$ is required for a Kolmogorov-type diffusion and harder than $p=-2.3$ for an energy-independent diffusion, as shown in the left panels and right panels of Fig.\ref{fig:PSR B0656+14}, respectively. Similar to the case of Geminga, given a Kolmogorov-type diffusion for the inner region, the positron flux produced by PSR~B0656+14 contribute to  $\leq 15\%$ of the observed flux at energies below 400 GeV, with the spectrum rising too steeply compared with the data. However, if an energy-independent diffusion is assumed, PSR~B0656+14 is allowed to contribute up to 30\% of the observed positron flux at 100\,GeV, and up to 50\% of the positron flux at 400\,GeV by the current data.

\section{Summary}
{To summarize,  we have shown that \textsl{Fermi}-LAT's observations of the TeV nebulae of Geminga and PSR~B0656+14 provide more direct constraints on the positron density at GeV energies in the vicinity of the pulsars than the TeV observations, and hence on their contribution to the measured positron excess at Earth. We conclude that two nearby pulsars are disfavored as the dominant source of the positron excess at $\sim 50-500$ GeV in the two-zone diffusion models with the usual Kolmogorov-type diffusion coefficient. In this case, the positron excess must be largely contributed by other sources, such as other  pulsars, other types of nearby cosmic accelerators such as supernova remnants \citep{Fujita2009} and microquasars \citep{Gupta2014}, or even the annihilation or decay of dark matter particles(e.g. \citet{Cholis2013,Silk2014,Bergstrom2009,Yin2009}). However, for an energy-independent diffusion  in the TeV nebulae, a dominant contribution of the positron excess  by them cannot be ruled out by the current data.

This work has made use of data and software provided
by the Fermi Science Support Center.
X.Y.W. is supported by the National Key R \& D
program of China under the grant 2018YFA0404200 and the NSFC  grants
11625312 and 11851304.

\begin{table*}
\centering
\begin{threeparttable}
\caption{Statistic test and $95\%$ upper limit fluxes for each extended Template}
\begin{tabular}{lcc|ccc}
\hline
\hline
Model name  & UL (10-500 GeV)  &TS& Model name  &UL (10-500 GeV)  &TS  \\
(Geminga) & $\rm \times 10^{-10} ph~cm^{-2}~s^{-1}$ &&(PSR~B0656+14)&$\rm \times 10^{-10} ph~cm^{-2}~s^{-1}$  &\\

\hline
$2^{\circ}$ Disk & 0.44 & 0.1 &$2^{\circ}$ Disk&  0.61& 0.3 \\
Diffusion 1 & 29.1 & 16.1 & Diffusion 6 & 2.8 & 0.5 \\
Diffusion 2 & 6.3 & 4.1 & Diffusion 7 & 4.9 & 0.5 \\
Diffusion 3 & 4.5 & 2.1&  &  &  \\
Diffusion 4 & 14.3 &  6.6&  &  &   \\
Diffusion 5 & 3.9 & 1.0 &  &  &   \\

\hline
\hline
\end{tabular}
\begin{tablenotes}
    \item{Notes. The upper limit fluxes in 10-500 GeV at $95\%$ confident level. The meaning of the model names for Geminga are as follows: 1) Disk represents an uniform disk with a radius of  $2^\circ$ similar to that in the HAWC catalog \citep{Abeysekara2017a}; 2) Diffusion 1 template corresponds to the $\textsl{Fermi}-LAT's$ gamma-ray profiles predicted by the one-zone model adopting the same parameters as in ref. \cite{Abeysekara2017} (see the left panel of Fig 1);  3) Diffusion 2-5  templates correspond to two-zone models with the parameters the same as that used in the Fig.3 to Fig.7, respectively. The meaning of the model names for PSR~B0656+14 are as follows:  1) Disk represents an uniform disk with a radius of $2^\circ$ similar to that in the HAWC catalog\citep{Abeysekara2017a}; 2) The Diffusion 6 and Diffusion 7 templates correspond to the two-zone  Diffusion model for a Kolmogorov-type diffusion and an energy-independent diffusion in the inner region, respectively. Note that all the Diffusion templates are cut at $25^\circ$. }
\end{tablenotes}
\label{tab1}
\end{threeparttable}
\end{table*}

\begin{figure}[htbp]
\centering
\includegraphics[width=0.4\textwidth]{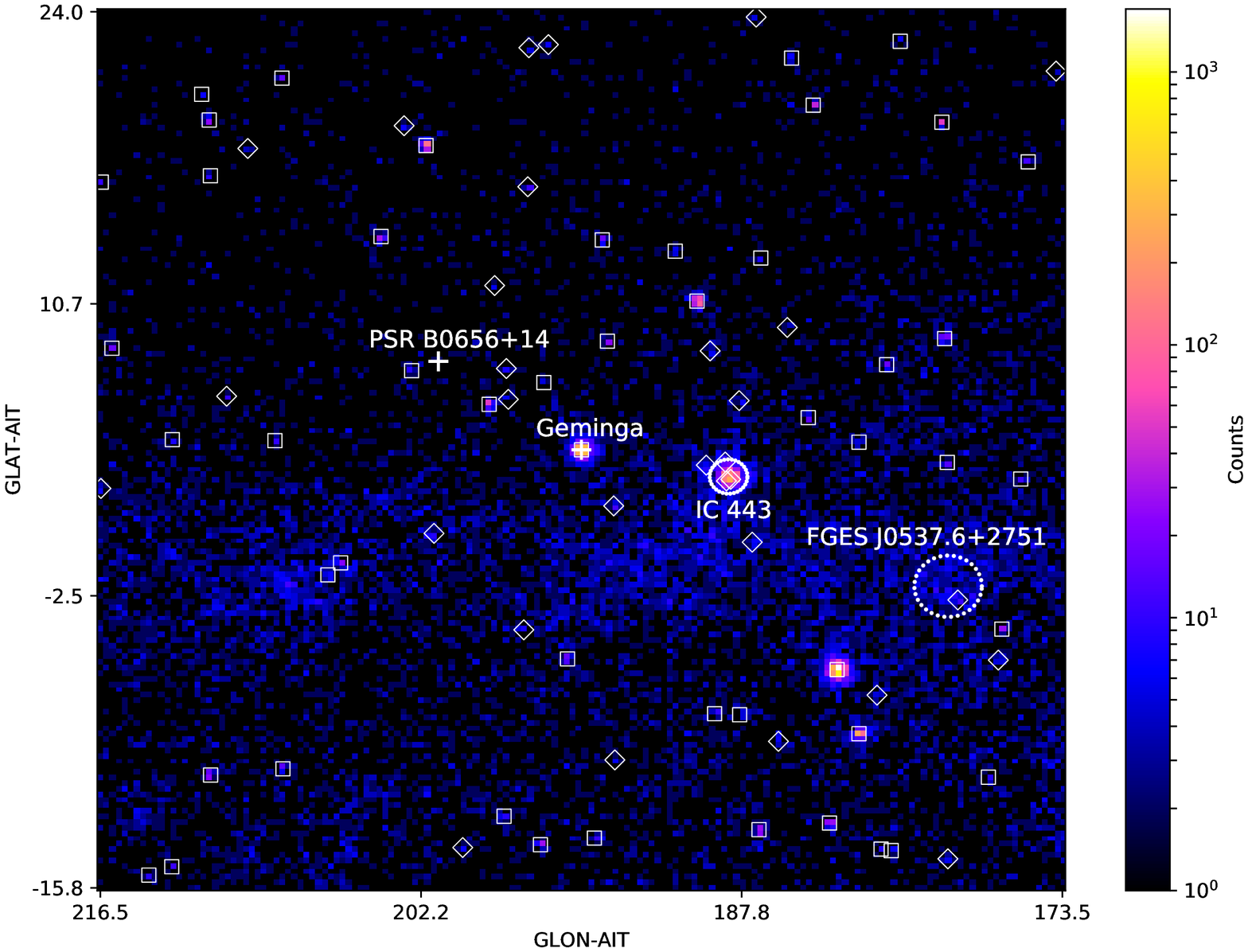}
\includegraphics[width=0.4\textwidth]{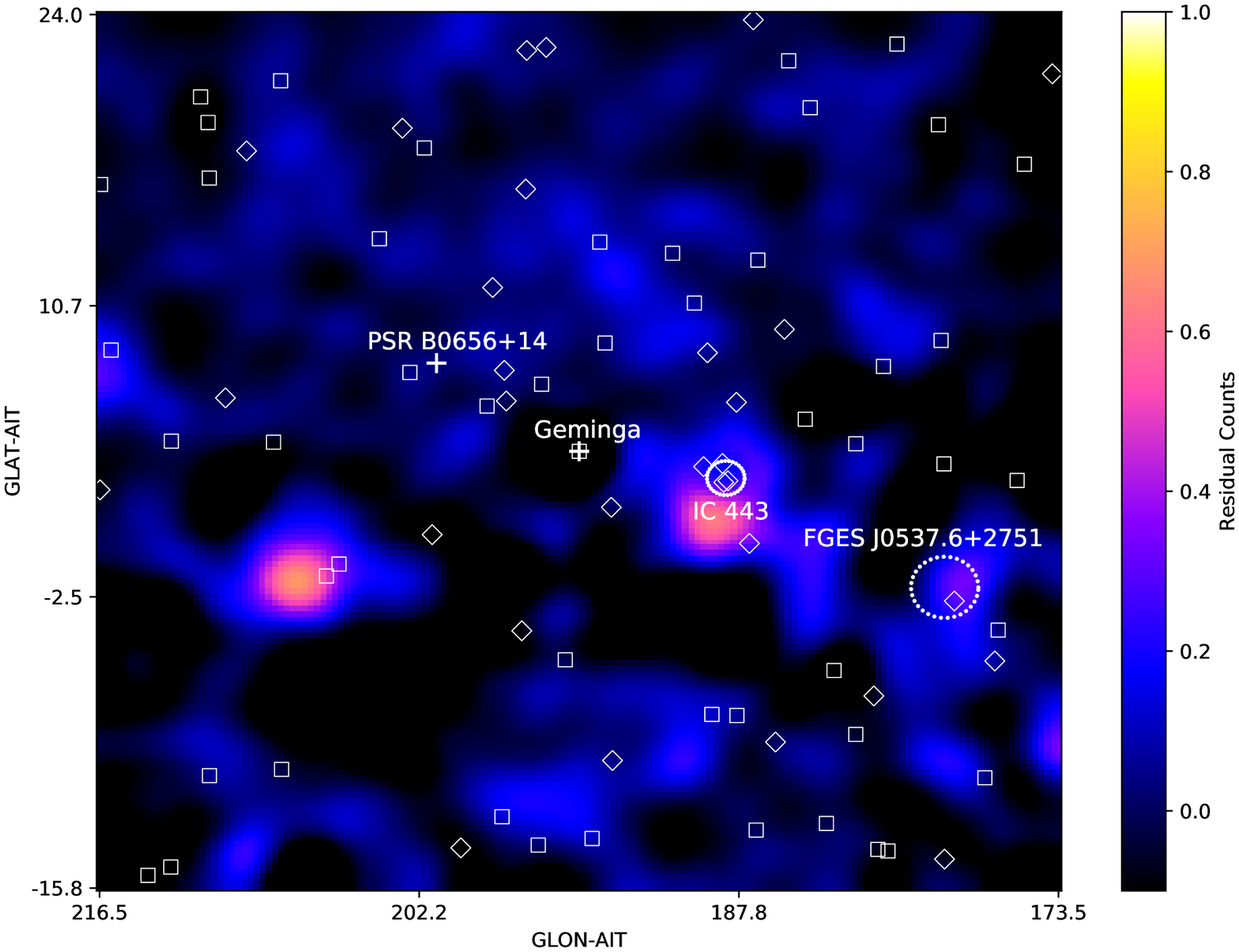}
\vspace{40pt}
    \caption {Gamma-ray count map (left) and residual maps subtracted all background components (right) for the $40^{\circ}\times40^{\circ}$ region in the energy band $10 - 500$ GeV. The crosses are located at the positions of Geminga and PSR B0656+14 respectively. The squares represent the 3FHL point sources.  The dotted circles are the extended source IC 443 modelled by a $0.78^{\circ}$ ($3\sigma$) Gaussian template and the extended source FGES J0537.6+2751 modelled by a $1.394^{\circ}$ disk template respectively.The diamonds represent the non-3FHL sources identified. Events were spatially binned in the region of side length $0.25^\circ$. The residual map is smoothed by gaussian kernel ($\sigma = 1^\circ$).}
\label{fig_cs}
\end{figure}

\begin{figure}
\centering
\includegraphics[width=0.45\textwidth]{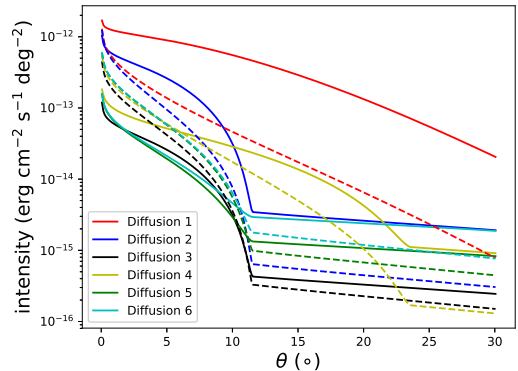}
\caption{  The surface brightness profiles at 10 GeV (solid line) and at 500 GeV (dashed line) for each diffusion models listed in Table I.}\label{fig:SBPGeV}
\end{figure}
\begin{figure}
\centering
\includegraphics[width=0.45\textwidth]{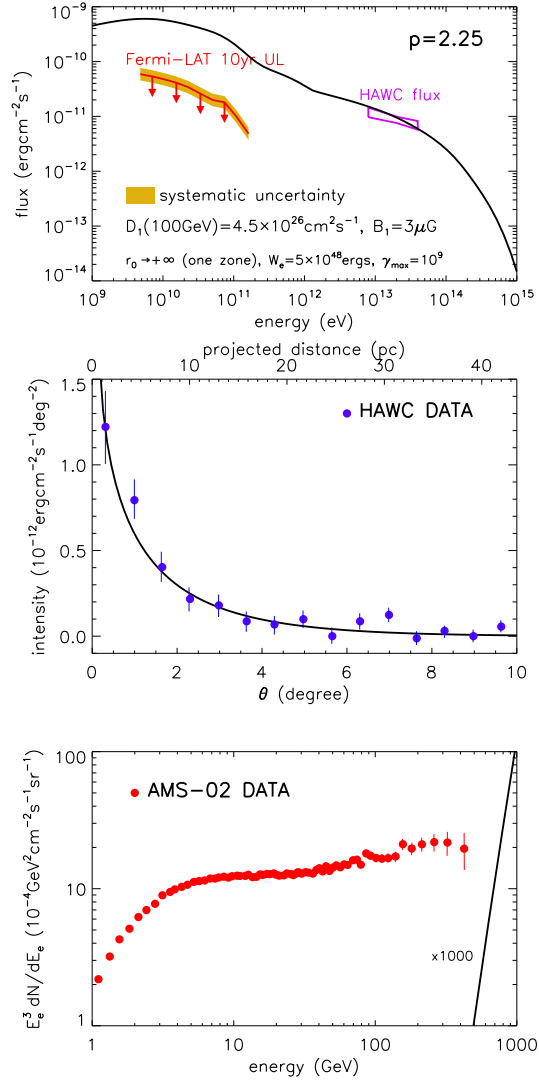}
\caption{Comparison between observational data and theoretical expectation in the one-zone diffusion model for the gamma-ray spectrum within $25^\circ$ of Geminga (top panel), the surface brightness profile of $8-40\,$ TeV emission  (middle panel) and  the positron flux at Earth (bottom panel). The GeV flux limits  are obtained using the diffusion spatial templates in accordance with the respective theoretical models (see Table 1). The TeV spectral and SBP data measured by HAWC are taken from ref.\cite{Abeysekara2017}. The positron flux data measured by AMS-02 are taken from \cite{Aguilar14}.}\label{fig:fitting_hawc}
\end{figure}

\begin{figure}
\centering
\includegraphics[width=0.45\textwidth]{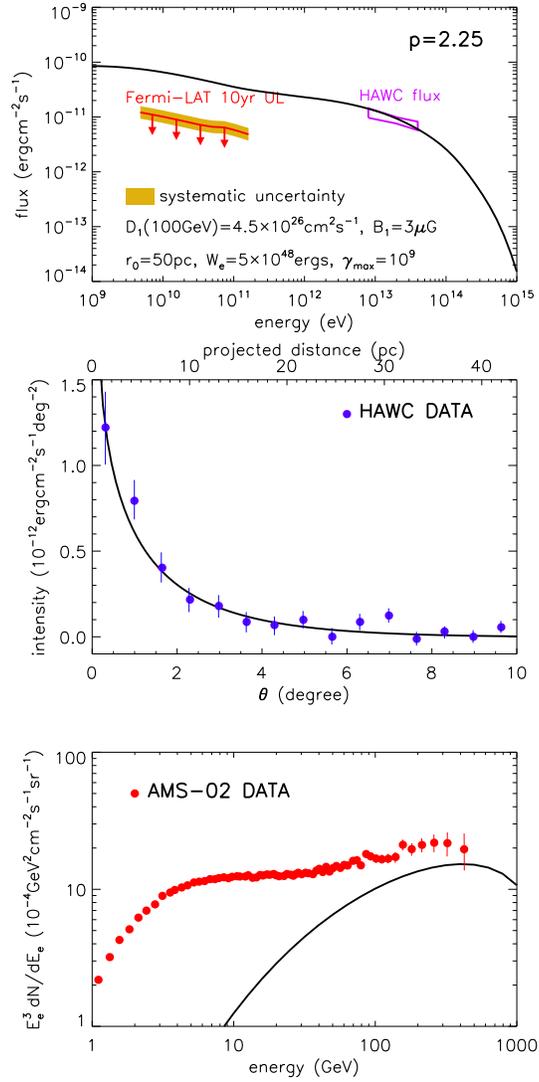}
\caption{Comparison between observational data and theoretical expectation in the  two-zone diffusion model where a standard ISM diffusion is assumed for the outer zone beyond $r_0=50 {\rm pc}$. }\label{fig:Diffusion2}
\end{figure}

\begin{figure}
\centering
\includegraphics[width=0.45\textwidth]{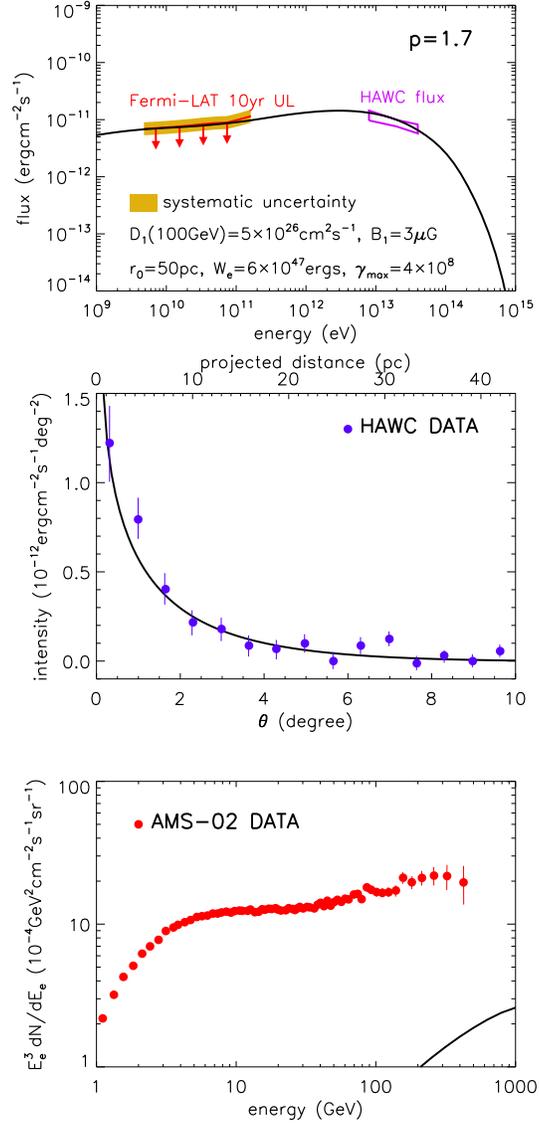}
\caption{Same as Fig.~\ref{fig:Diffusion2}, but considering the constraint from the \textsl{Fermi}-LAT's observation. }\label{fig:2D}
\end{figure}

\begin{figure}
\centering
\includegraphics[width=0.45\textwidth]{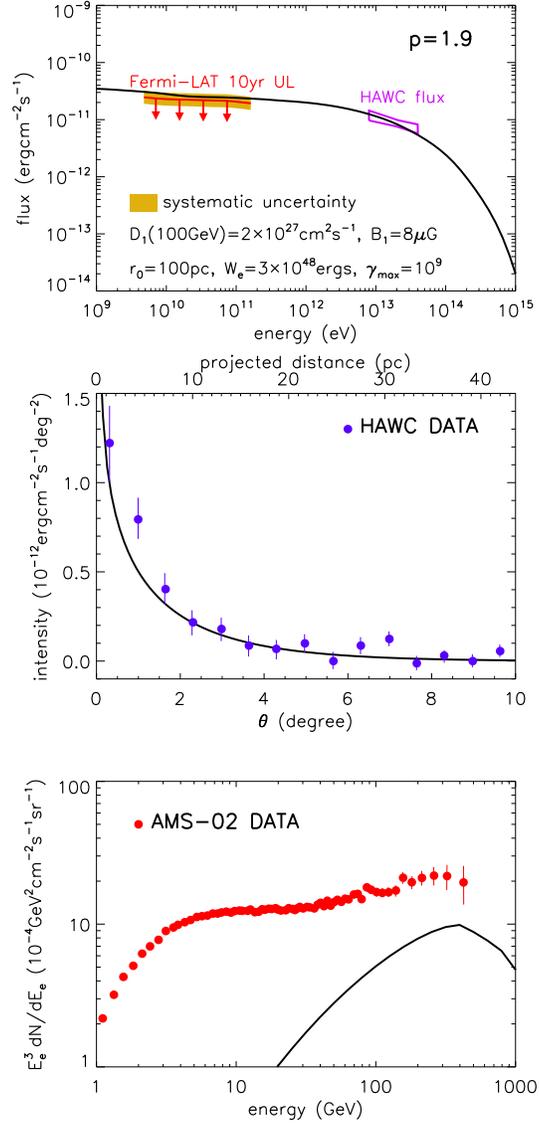}
\caption{Same as Fig.~\ref{fig:2D}, but assuming a larger boundary radius of $r_0=100 {\rm pc}$.}\label{fig:2D-100pc}
\end{figure}

\begin{figure}
\centering
\includegraphics[width=0.45\textwidth]{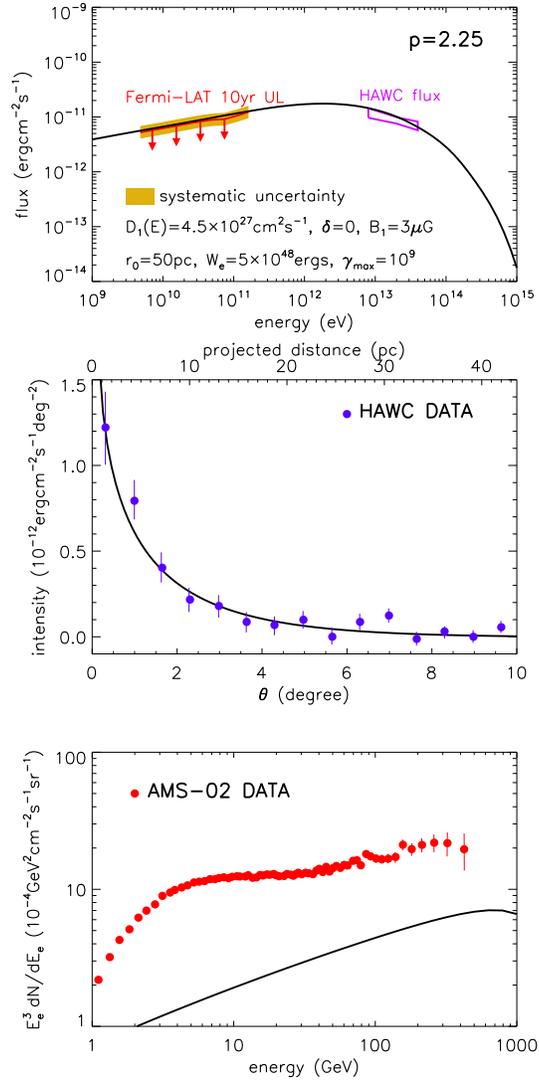}
\caption{Results for the case assuming an energy-independent diffusion in the inner region around Geminga. }\label{fig:constant-D1}
\end{figure}

\begin{figure}[htbp]
\centering
\includegraphics[width=0.45\textwidth]{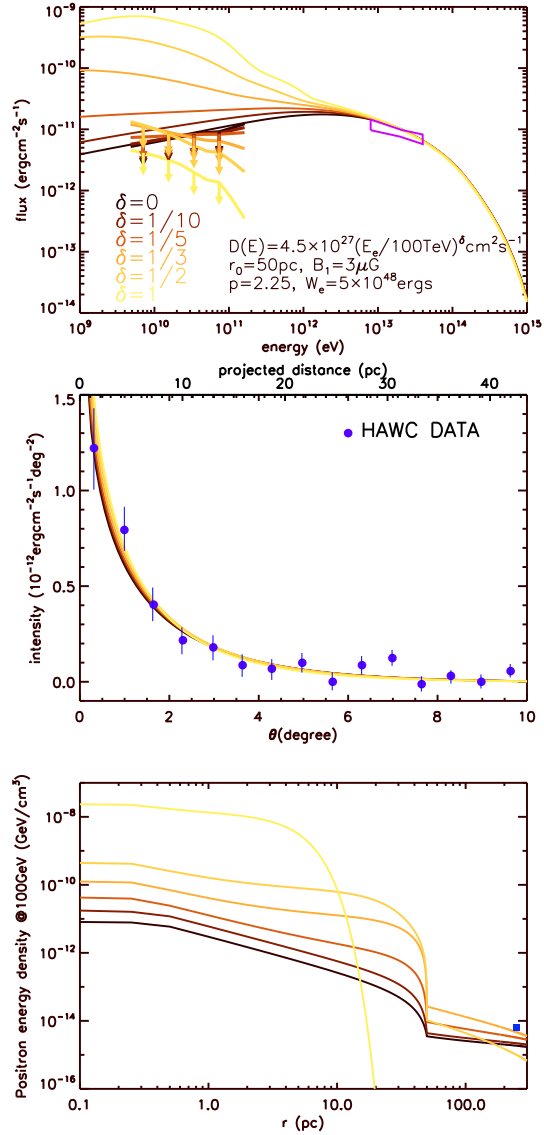}
\caption{Results for the different energy dependence ($\delta$) of the diffusion coefficient in the inner zone $D_1$. The diffusion coefficient is fixed to be $4.5\times 10^{27}\rm cm^2s^{-1}$ at 100\,TeV, following the best-fit value obtained in HAWC's paper \cite{Abeysekara2017}. Other parameters are the same as those in the benchmark case. {\bf Top panel}: gamma-ray flux from the region of a radius of $25^\circ$ centered at Geminga; {\bf Middle panel}: $8-40\,$TeV SBP; {\bf Bottom panel}: The positron energy density at 100\,GeV as a function of the distance to Geminga. Earth is located at $r=250\,$pc. The blue square shows the 100\,GeV positron energy density at  Earth inferred from the AMS-02 measurements.}\label{fig:delta}
\end{figure}

\begin{figure}[htbp]
\centering
\includegraphics[width=0.45\textwidth]{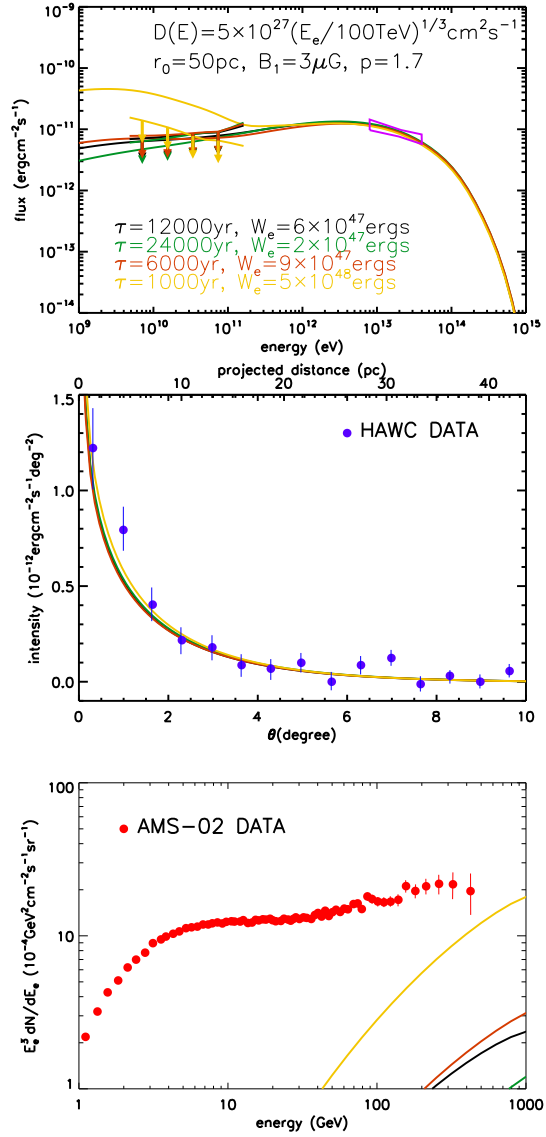}
\caption{The effect of  the spin-down timescale $\tau$. The employed parameters are labeled in the figure. See section~\ref{sec:tau} for more details.}\label{fig:tau}
\end{figure}
}

\begin{figure}[htbp]
\centering
\includegraphics[width=0.45\textwidth]{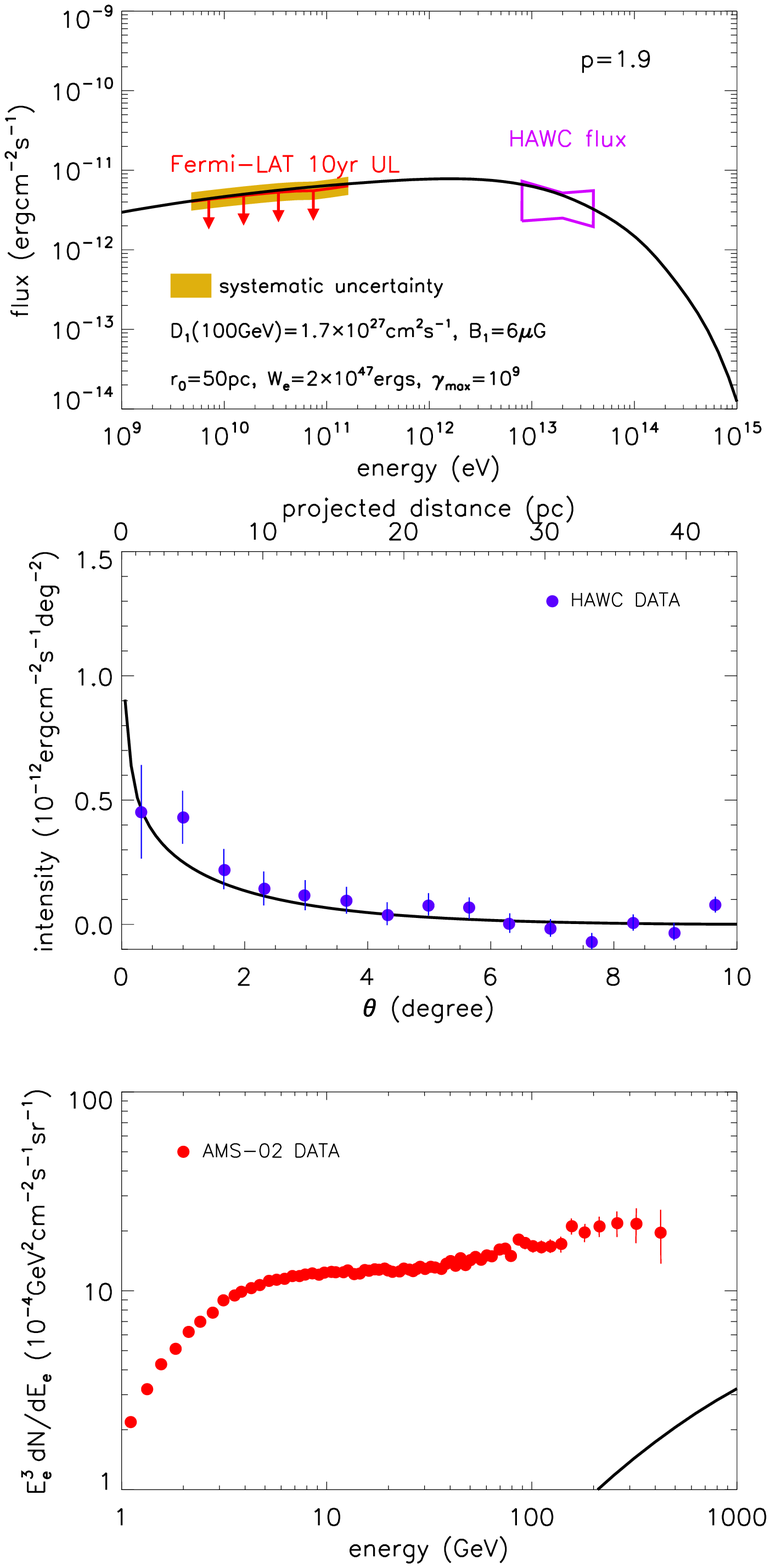}
\includegraphics[width=0.45\textwidth]{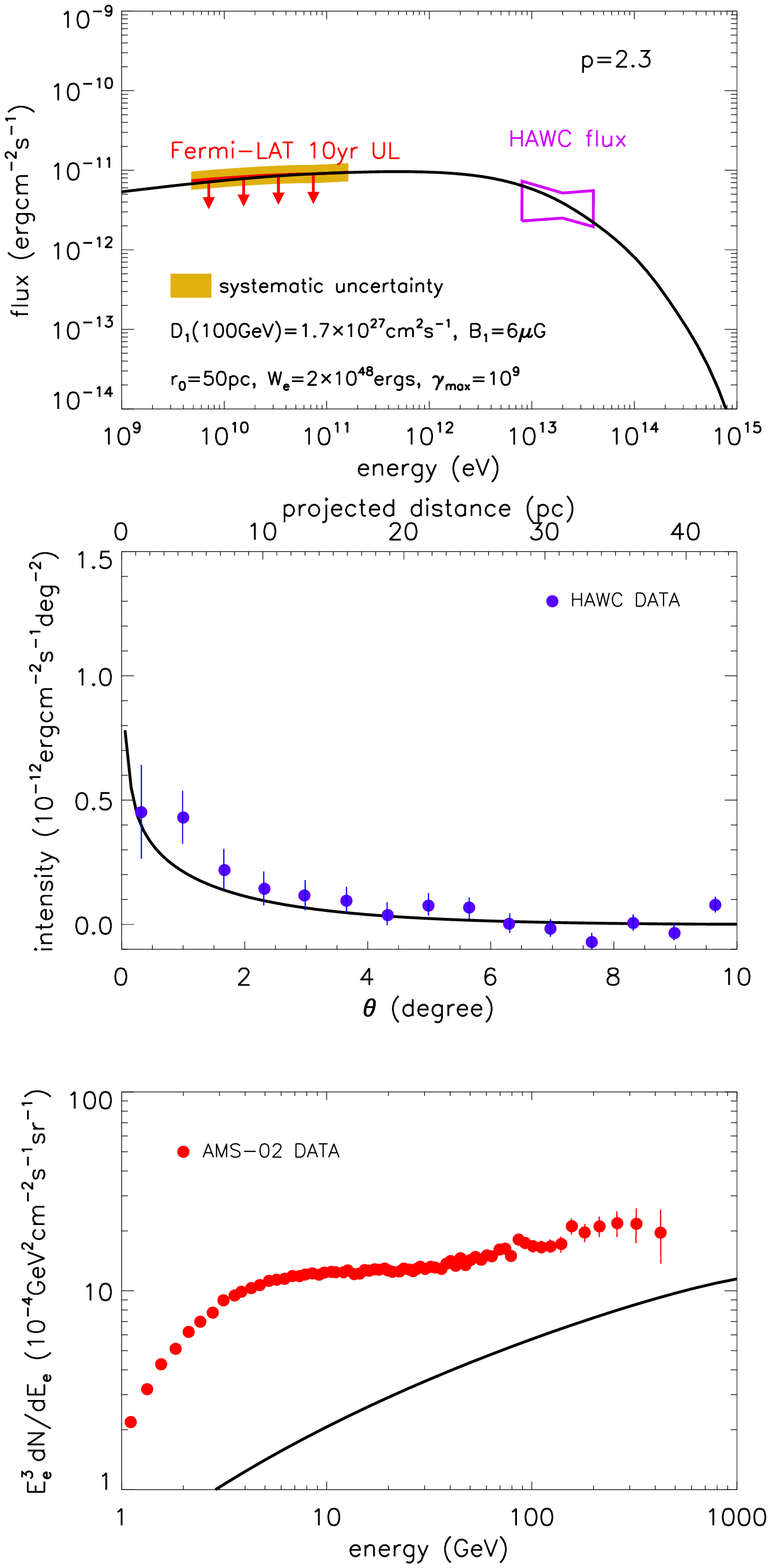}
\caption{Top panel: comparison between the expected flux and the observation within $25^\circ$ of PSR~B0656+14. Middle panel: the fit  of the SBP data of PSR~B0656+14 measured by HAWC. Bottom panel: comparison between the expected positron flux produced by PSR~B0656+14 and the observed flux at Earth. Left panels are under the assumption of a Kolmogorov-type diffusion while right panels are under the assumption of an energy-independent diffusion.}\label{fig:PSR B0656+14}
\end{figure}

\clearpage

\appendix
\section{Appendix:  Systematic uncertainty}
To assess the robustness of our results listed in Table I of the main text, we perform a number of systematic checks. Taking two templates (the Diffusion 1 and Diffusion 2 templates) as examples, we quantify the variation of the upper limits and the TS values. We first investigate the uncertainty from  Galactic interstellar diffuse emission (IEM) background modelling.  IEM is produced through interactions of high energy CR hadrons and leptons with interstellar gas via nucleon-nucleon inelastic collisions and electron bremsstrahlung, and with low energy radiation fields via IC scattering. The standard IEM template released by the \textsl{Fermi}-LAT collaboration is created using the assumption that energetic cosmic rays uniformly penetrate all gas phases of the interstellar medium. Under this assumption, the IEM intensities at any energy can be modelled as a linear combination of templates of gas column densities, an IC intensity map, the templates that partially accounts for the emission from Loop I and the \textsl{Fermi} bubbles. The linear combination coefficients can be determined from a fit to gamma-ray data \citep{2016ApJS..224....8A}. To explore the uncertainties related to this standard modelling of interstellar emission, we generated eight alternative IEMs to probe key sources of systematics by varying a few important input parameters to GALPROP (see Table~\ref{tab_IEM}). In addition, we adopt a different model building strategy from the standard IEM.  Starting from the work in ref. \cite{2012ApJ...750....3A} which provides 128 sets of intensity map associated with $\rm H_I$,$\rm H_{II}$, CO, and IC, we select eight intensity map sets corresponding to the model parameters listed in Table~\ref{tab_IEM}. The intensity maps associated with gas are binned into four Galactocentric rings ( 0-4 kpc, 4-8 kpc, 8-10 kpc and 10-20 kpc ).  We fit all model intensity maps simultaneously with an isotropic component, Loop I, Fermi bubbles, and 3FGL sources \citep{2015ApJS..218...23A} to 8 years of \textsl{Fermi}-LAT data in order to determine the linear combination coefficients. We use eight alternative IEM templates to replace the standard one and perform the likelihood analysis to estimate uncertainties {in the upper limits and TS values}. We emphasize that  these eight templates do not reflect the complete uncertainty of the IEMs, but the resulting uncertainty should be a reasonable indicator of the systematic error due to the mismodelling of the IEM background. We find that varying the IEM models increases the upper limits by at most $23\%$, as shown in Fig. \ref{fig:gll}.

Moreover, we study the systematic uncertainty caused by different selections of the ROI. Using the $40^\circ \times 40^\circ$ square ROI paralleling to right ascension and declination (dubbed as  alt. ROI), we find that the TS value of the Diffusion 1 template decreases significantly and the upper limits also decrease by $\sim 20\%$.  We further check the results using a smaller ROI ($30^\circ \times 30^\circ$) and a larger ROI ($50^\circ \times 50^\circ$) respectively, and find that the results derived from the smaller ROI are close to that of alt.ROI analysis and the results from the larger ROI are close to that of the standard analysis. We also test the systematic uncertainty using different event class data, which can be considered as an indicator for checking the impact of isotropy emission component. For the upper limit estimation, the systematic uncertainty caused by the uncertainty in the effective area is also included.

We summarize the above systematic uncertainties in Table~\ref{tab_un}.   {It is worth  noting that  the TS values of the Diffusion 1 template ranges from 5 to 33 in our systematic checks,  which may indicate that the detection significance suffers from systematic uncertainties.  To be conservative, we here give the upper limits on the GeV flux,  which can used to constrain the contribution to the positron flux by Geminga. } {The maximum of the total systematic uncertainty for the Diffusion 2 template is  $32\%$. For simplicity, we use this maximum value as an approximation for the total systematic uncertainty of all other diffusion templates as well.}

\begin{table*}
\centering
\begin{threeparttable}
\caption{ Systematic Uncertainties}
\begin{tabular}{lcccccc}
\hline
\hline
Type & Variation of       &UL Impact  &TS                 &UL Impact& TS     \\
       & Input Parameters& (diffuse 1)         &(diffuse 1) & (diffuse 2)     &  (diffuse 2) \\
\hline

ROI$^{\rm a}$   & $30^\circ$& $ -33\%$ & $ 2.5$ & -24\% & 2.3 \\
                         & $50^\circ$ & $ 6\%$ &  32.7  &  $16\%$ & 7.7 \\
               & alt. $40^\circ$  & $ -34\%$ &  5.8  &  $-18\%$ & 1.8 \\
Event class$^{\rm b}$ &alt. Event classes & $ (-10\%,6\%)$ &  (8.7,18.9)  &  $(-1\%,12\%)$ & (2.8,7.7) \\
Diffuse modelling $^{c}$& alt. diffuse models & $ (-23\%,21\%)$ &$(4.7,26.9)$& $(-26\%,23\%)$& (1.1, 8.6)\\
Effective Area$^{\rm d}$ &  & $(-10\%,10\%)$  & & $(-10\%,10\%)$& \\
Total$^{\rm e}$ &  & $(-43\%,25\%)$  & &$(-37\%,32\%)$& \\
\hline
\hline
\end{tabular}
\begin{tablenotes}
\item\textbf{Notes.} Overview of systematic uncertainties.
\item[a]  We select the ROI of $30^\circ \times 30^\circ$, $50^\circ \times 50^\circ$ and alternative $40^\circ \times 40^\circ$  centered at the position of Geminga to repeat our analysis. The alt. $40^\circ$  represents $40^\circ \times 40^\circ$ square ROI that is parallel to right ascension and declination. {Note that we get a lower TS value, $\sim 22$, in the analysis of ROI $50^\circ \times 50^\circ$ using the newest released IEM ($\rm gll\_iem\_v07.fits$) and the background sources of the Fourth Fermi-LAT Source Catalog (4FGL)}.
\item[b] The CLEAN, UTRACLEAN, UTRACLEANVETO and SOURCEVETO conversion  events are used to perform the analysis..
\item[c] See the Table~\ref{tab_IEM}  for details.
\item[d] Generally, the uncertainties in the effective area are estimated by using modified IRFs whose effective area bracket that of our nominal IRF. These "biased" IRFs are defined by envelopes above and below the nominal dependence of the effective area with energy by  linearly connecting differences of (5\%, 10\%) at log(E)  of (4,  5.2) respectively \citep{2012ApJS..203....4A}. We simply  consider the uncertainties of effective area for estimating upper limit as $\sim 10\%$ in our analysis energy band.
\item[e] We combine all the errors in quadrature to obtain our total systematic uncertainty.

\end{tablenotes}
\label{tab_un}
\end{threeparttable}
\end{table*}

\begin{table*}
\centering
\begin{threeparttable}
\caption{Alternative Models for Galactic Diffuse Emission}
\begin{tabular}{lccccc}
\hline
\hline
Model number  & CR Source Distribution  &Halo Size& Spin Temperature  \\
&  &(kpc)  &(K)\\

\hline
1 &  Lorimer & 4 & 150 \\
2 &  Lorimer & 4 & $10^5$ \\
3 &  Lorimer & 10 & 150 \\
4 &  Lorimer & 10 & $10^5$ \\
5 &  SNR & 4 & 150 \\
6 &  SNR & 4 & $10^5$ \\
7 &  SNR & 10 & 150 \\
8 &  SNR & 10 & $10^5$ \\

\hline
\hline
\end{tabular}
\begin{tablenotes}
    \item{Notes. Overview of the alternative IEM models used for assessing the systematic uncertainties. We chose to vary the three most important input parameters that were found in scanning the parameter space in ref. \cite{2012ApJ...750....3A}. Note that we use radial boundaries of 20 kpc and apply a magnitude cut to the E(B-V) maps  at 5. For the CR source distribution, we adopted the distribution of SNRs according to \cite{1998ApJ...504..761C} and of pulsars according to \cite{2006MNRAS.372..777L};
}
\end{tablenotes}
\label{tab_IEM}
\end{threeparttable}
\end{table*}

\begin{figure}[htbp]
\centering
\includegraphics[width=0.4\textwidth]{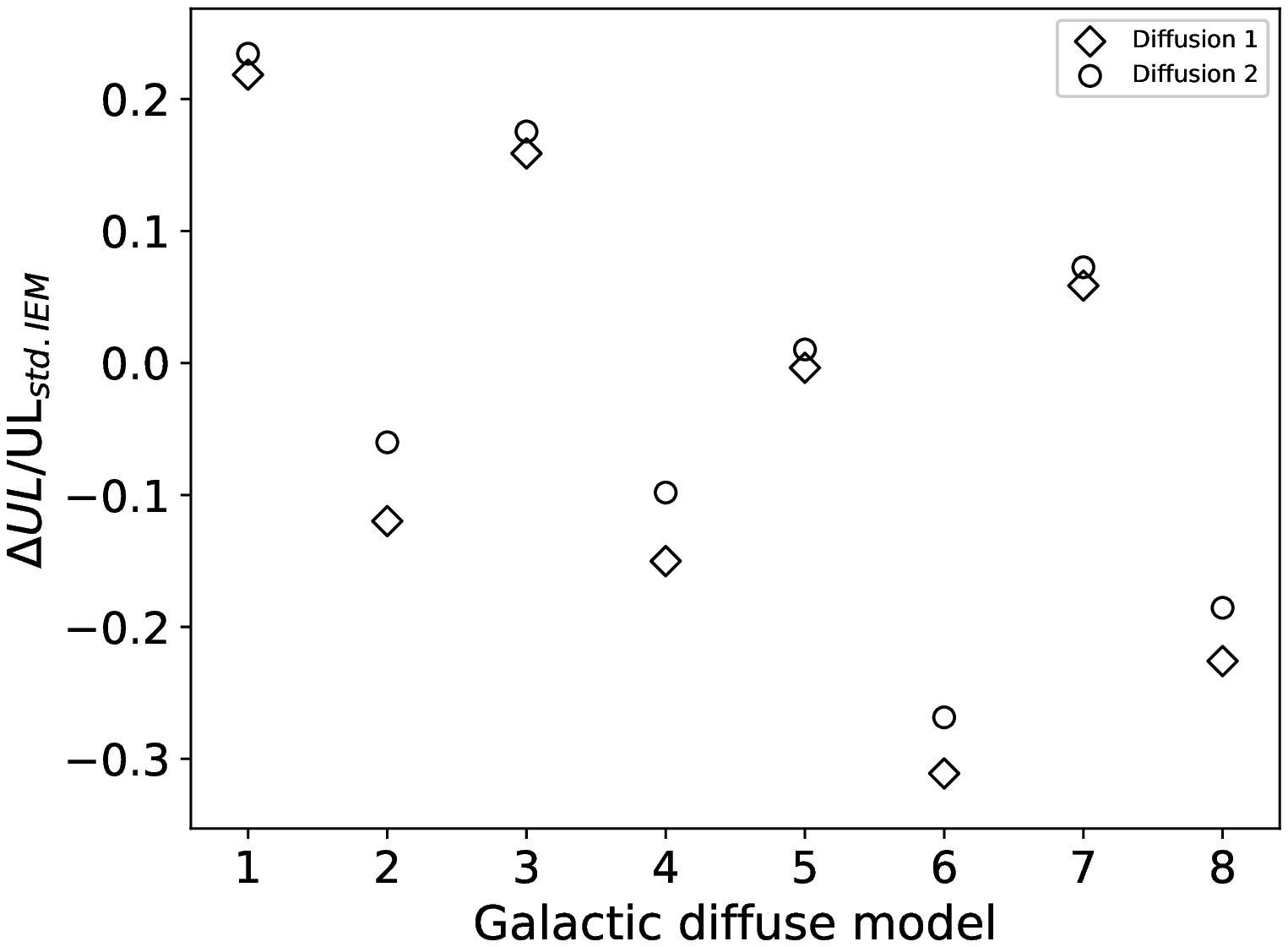}
\includegraphics[width=0.4\textwidth]{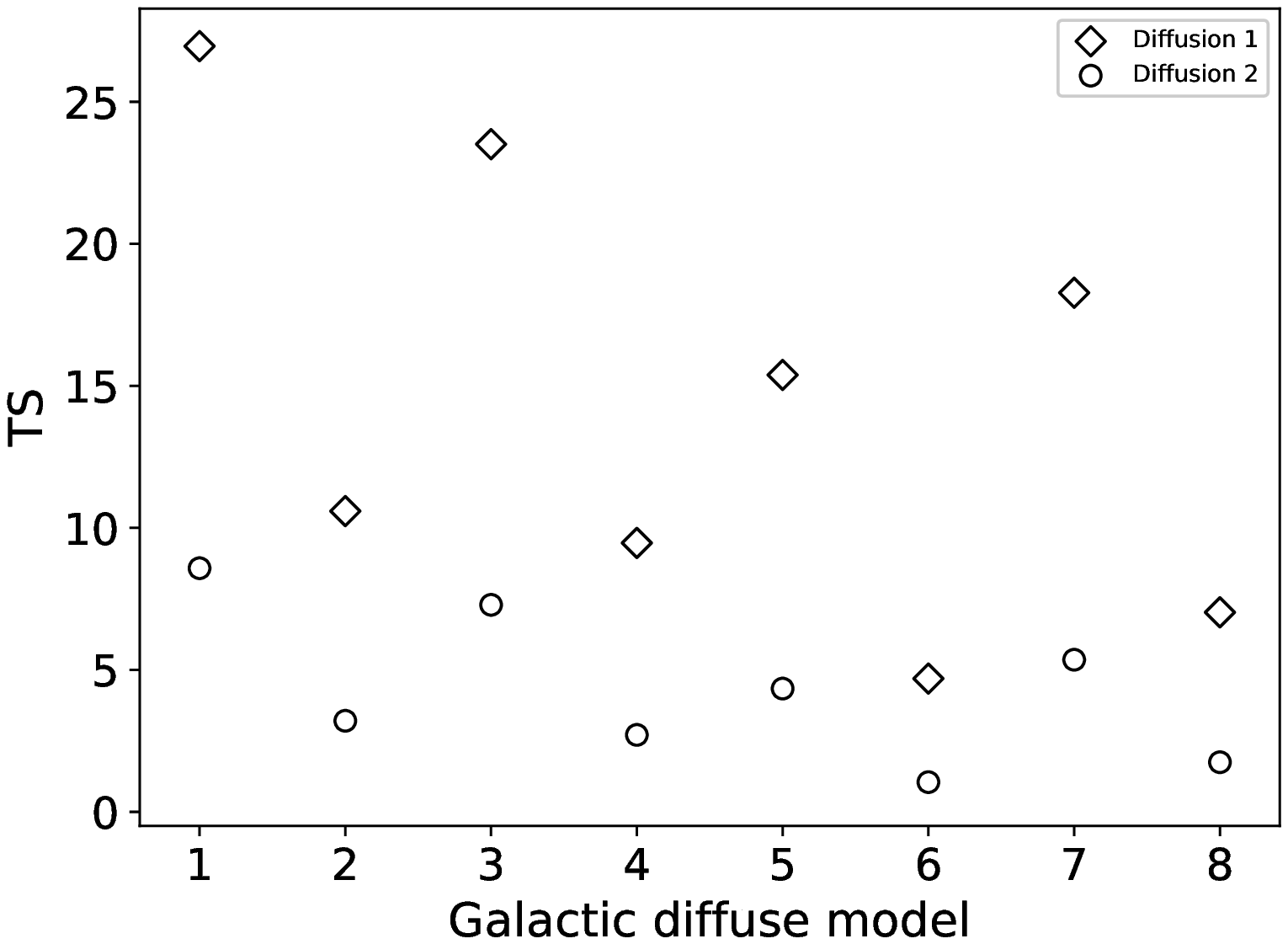}
\vspace{40pt}
    \caption {Left: the $95\%$
 upper limit variations of the Diffusion 1 and Diffusion 2 for alternative IEM templates  compared to that for the standard IEM template. Right: the TS values for the Diffusion 1 and Diffusion 2 templates for each alternative IEM templates.}
\label{fig:gll}
\end{figure}

\end{document}